\documentclass{moriond}

\usepackage{xspace}
\usepackage{amsmath}

\newcommand{\Photo}{}

\newcommand{\nue}{$\nu_e$\xspace}
\newcommand{\nuecc}{$\nu_e~\text{CC}$\xspace}
\newcommand{\nueccnopinp}{$\nu_e~\text{CC} 0\pi \text{N}p$\xspace}
\newcommand{\nueccnopinop}{$\nu_e~\text{CC} 0\pi 0p$\xspace}
\newcommand{\nueccmpinp}{$\nu_e~\text{CC} \text{M}\pi \text{N}p$\xspace}

\newcommand{\numucc}{$\nu_{\mu}~\text{CC}$\xspace}

\let\OLDthebibliography\thebibliography
\renewcommand\thebibliography[1]{
  \OLDthebibliography{#1}
  \setlength{\parskip}{0pt}
  \setlength{\itemsep}{0pt plus 0.3ex}
}

\begin{document}
\vspace*{4cm}
\title{SEARCH FOR A LOW ENERGY EXCESS IN MICROBOONE}

\author{NICOL\`O FOPPIANI \\ on behalf of the MicroBooNE collaboration}

\address{Department of Physics, Harvard University, Cambridge, MA, 02138}

\maketitle

\abstracts{
MicroBooNE (the Micro Booster Neutrino Experiment) is a liquid argon time-projection chamber (TPC) experiment designed for short-baseline neutrino physics, currently running at Fermilab.
It aims to address the anomalous excess of low-energy events observed by the MiniBooNE experiment.
Recent progress towards the search for the low-energy events have brought to develop fully automated event selection algorithm to identify charged-current electron neutrino event candidates with no pions and at least one proton in the final state (\nueccnopinp) using the Pandora multi-algorithm pattern recognition.
Several cross checks and sidebands have been studied so far to validate the analysis.
}

\section{The Low Energy Excess}

\begin{figure}
\begin{minipage}{0.5\columnwidth}
\centerline{\includegraphics[width=0.9\columnwidth]{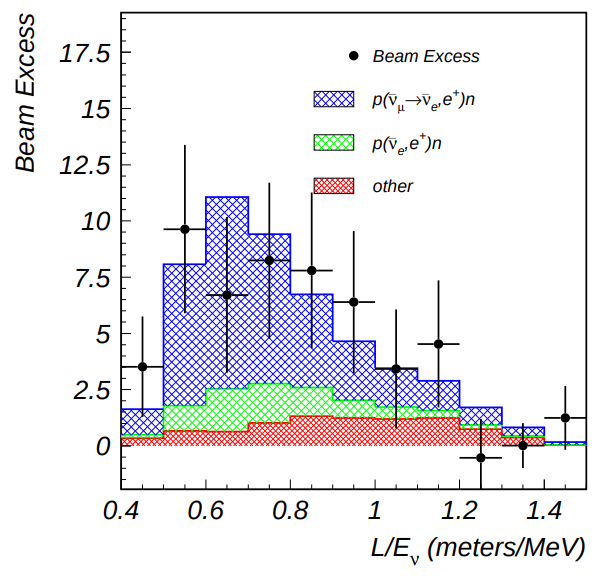}}
\end{minipage}
\hfill
\begin{minipage}{0.5\columnwidth}
\centerline{\includegraphics[width=0.9\columnwidth]{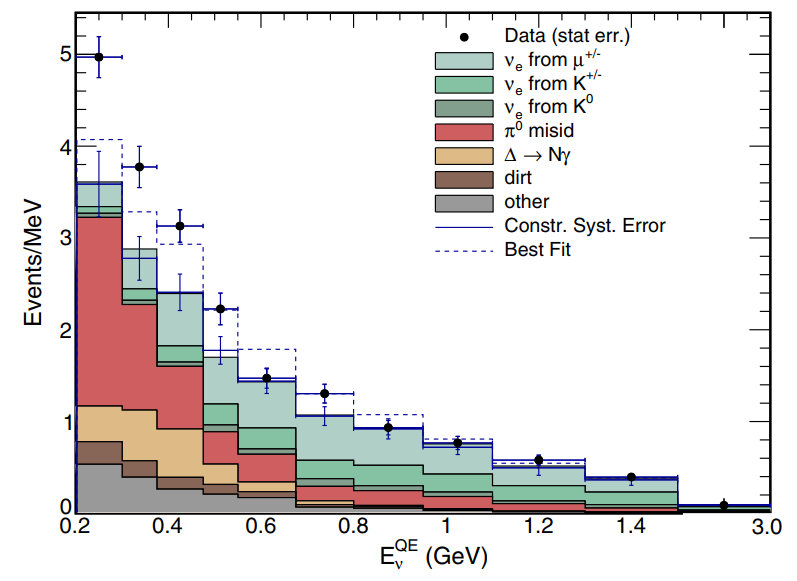}}
\end{minipage}
\caption[]{Energy distribution of the selected events by the LSND (left) and MiniBooNE (right) experiments, showing the excess of events with respect to the prediction. The MiniBooNE excess is often referred as the Low Energy Excess.}
\label{fig:01_lee}
\end{figure}

Several short baseline neutrino experiments have observed anomalies which can be explained with the presence of additional sterile neutrinos in the eV mass range.
The Liquid Scintillator Neutrino Detector (LSND) and the MiniBooNE experiments observed excesses of electromagnetic-like events with respect to the predictions in the low energy regions \cite{lsnd} and \cite{miniboone}.
The two plots in figure \ref{fig:01_lee} show the comparisons between the data and the simulation as observed by LSND (left) and MiniBooNE (right).
The excess, often referred as Low Energy Excess (LEE) has been interpreted as due to the presence of an additional sterile neutrino which amplifies the oscillation $\nu_{\mu} \rightarrow \nu_e$ through $\nu_{\mu} \rightarrow \nu_{sterile} \rightarrow \nu_e$.
The main goal of the The Micro Booster Neutrino Experiment (MicroBooNE) experiment is to investigate the origin of this anomalies.
MicroBooNE is a liquid argon time projection chamber, designed to measure neutrino interactions with bubble chamber resolution, providing capability to distinguish electromagnetic showers originated by electrons and photons.
For this reason MicroBooNE might be able to shed new light on the LEE.
MicroBooNE is located at Fermilab, close to MiniBooNE, in order to explore the same baseline and energy regime explored by MiniBooNE, but with a new technology.
In addition to the LEE program, MicroBooNE will also be able to deliver several neutrino-nucleus cross section measurements in liquid argon.
As liquid argon is the technology that has been chosen for the Deep Underground Neutrino Experiment (DUNE) \cite{dune}, these measurements will be extremely important in view of an accurate simulation of the expected event rates in DUNE.

\section{The MicroBooNE experiment}

\subsection{Beamlines}
Neutrinos detected by MicroBooNE come from two different beamlines.
BNB is the main beamline designed to achieve the physics goals of MiniBooNE and MicroBooNE.
% It will be eventually used for the whole Short Baseline Neutrino (SBN) program at Fermilab, which consists of three experiments, namely MicroBooNE, Short Baseline Neutrino Detector (SBND) and Icarus.
The second beamline exploited by MicroBooNE is NuMI, which is primarily designed and used by a different set of experiment, such as NOvA and MINERvA.
MicroBooNE detects off-axis neutrinos from NuMI: this implies larger uncertainties on the simulation of the neutrino flux with respect to BNB.
For this reason this second beamline is mainly used for cross checks and sidebands rather than to study oscillation physics.

\subsection{Detector and reconstruction}
\begin{figure}
\begin{minipage}{0.5\linewidth}
\centerline{\includegraphics[width=0.95\linewidth]{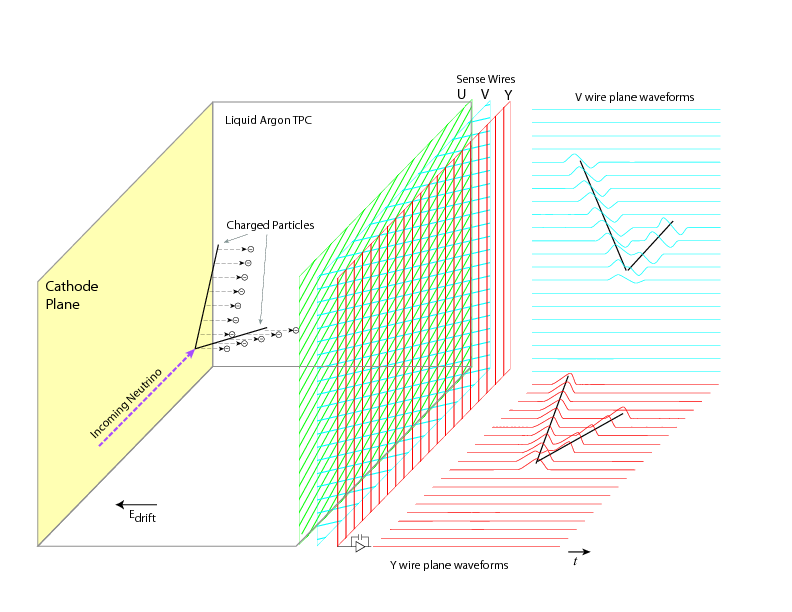}}
\end{minipage}
\hfill
\begin{minipage}{0.5\linewidth}
\centerline{\includegraphics[width=0.95\linewidth]{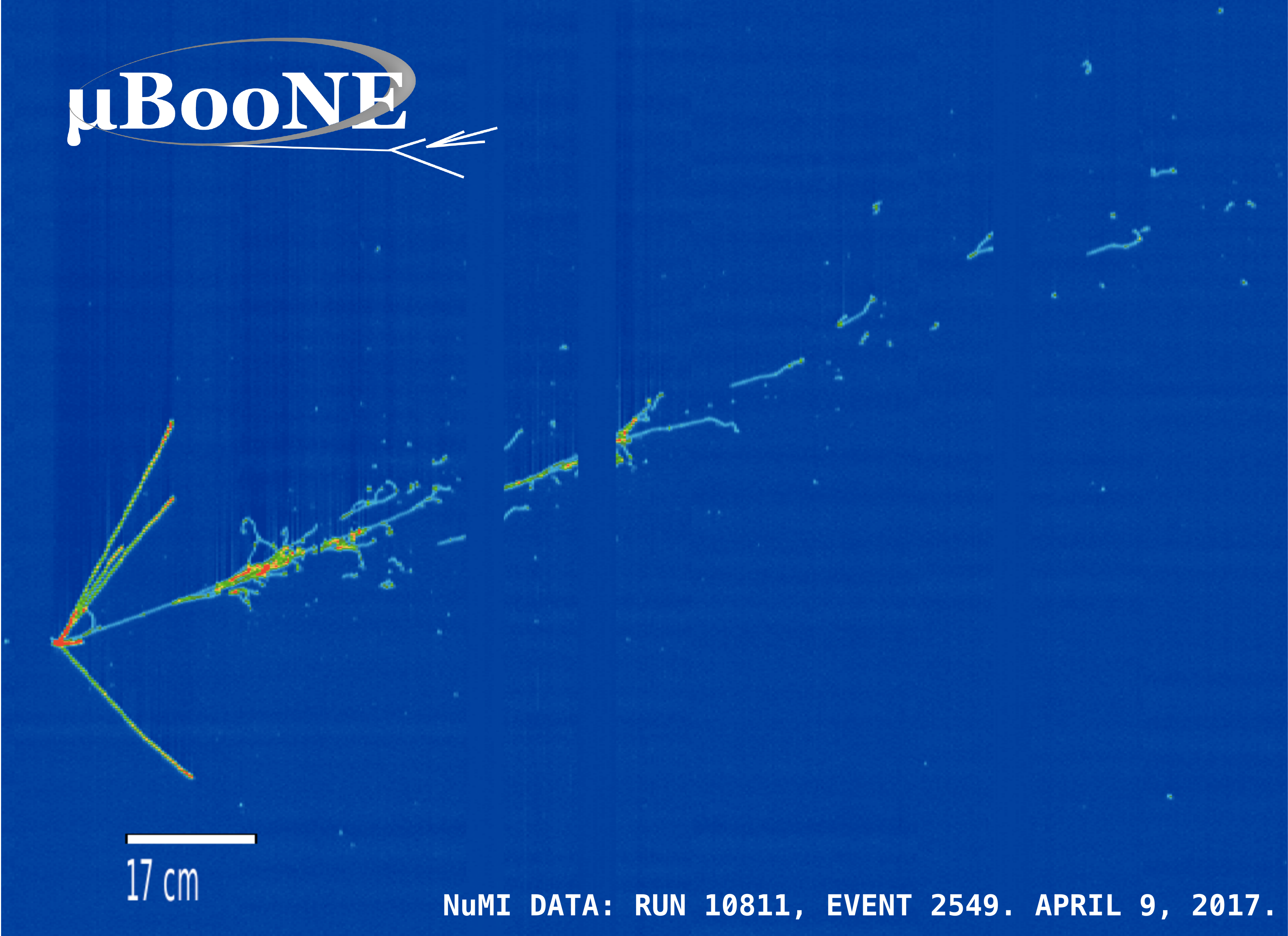}}
\end{minipage}
\caption[]{Left: Production of signals in a liquid argon TPC like MicroBooNE. Right: Event display of the charge deposited on the collection plane in a \nuecc interaction recorded in the NuMI dataset. The colormap shows the amount of deposited charge. Among the different particles associated with the interaction an electromagnetic shower generated by an electron is clearly distinguished.}
\label{fig:02_microboone}
\end{figure}

MicroBooNE \cite{microboone} is a liquid argon time projection chamber containing about 89 tons of liquid argon, in a volume of 10.4 (length) $\times$ 2.5 (width) $\times$ 2.3 (height) m.
Charged particles travelling through the detector ionise the argon, leaving a sea of electrons that are drifted in an orthogonal direction to the beam, through a 70kV electric field, as shown in the left drawing of figure \ref{fig:02_microboone}.
Because of the high purity of the argon, electrons can travel the whole detector with being little or no affected.
The ionisation is subsequently detected by three wire planes, providing three 2-dimensional images of the interactions.
An example of such an image for an event containing a \nue charged current interaction is shown in the right plot of figure \ref{fig:02_microboone}.
% Wires are spaced by 3mm, providing sub-cm spatial resolution.
The three 2-dimensional images are then combined together in order to produce a 3-dimensional reconstruction of the interaction, through the Pandora multi-algorithm pattern recognition software \cite{pandora}.
% The reconstruction is performed using the Pandora multi-algorithm pattern recognition \cite{pandora}.
% This software provides all the possible neutrino interactions, together with a list of reconstructed particles, tracks and showers, for each neutrino interaction.
% This is the basis of the analyses described in the following.

\section{LEE search}

The low energy excess search is structured as a blind analysis, developed on the Monte Carlo simulation.
% The data has not been unblinded yet, and it will only when the analysis is fully optimised and finalised.
Only small fraction of the data is currently available to evaluate the goodness of the simulation, and cross check the analysis using specific sidebands and control regions.
The datasets the results in this article are based on consist of $4 \times 10^{19}$ protons-on-target (POT) BNB (3.5\% of the total BNB data) and $2.4 \times 10^{20}$ POT NuMI (21\% of the total NuMI data).

\subsection{Categories and topologies}

There are two main searches currently ongoing within the MicroBooNE collaboration.
Firstly, a search for an electron-like excess.
% This analysis is also the main topic of the next sections of this article.
This channel tests the possibility that the low energy excess is produced through an enhanced oscillation $\nu_{\mu} \rightarrow \nu_e$.
This analysis aims to select electron neutrinos charged current interaction, characterised by the presence of an electromagnetic shower, which is attached to the main vertex, and whose energy deposition at the beginning of the shower is compatible with the value of a minimum ionising particle (mip) of about $dE/dx \simeq 2 \text{MeV}/\text{cm}$.
Within this search, several subchannels are considered.
The final state electron can be produced as the only particle in the final state (\nueccnopinop), in association with a certain number of protons (\nueccnopinp), or in association with protons and pions (\nueccmpinp).
The first channel is the most difficult to distinguish from the single photon production, whereas the third one typically consist of high energy interactions.
In order to investigate the LEE, the second channel is of primarily importance, as this was the category targeted by the MiniBooNE analysis that observed the LEE, and because the presence of additional protons allows an accurate reconstruction of the vertex, and a more precise tagging of the electron shower.
This subchannel is the main focus of the rest of the document.
% The channel in which the electron is the only final state particle produced in the interaction (\nueccnopinop).
% This channel has the advantage of being the most abundant, consisting of about half of the \nuecc interactions in the low energy regime.
% On the other hand, the photon background is very important for this channel, as no other vertex requirement can be used, as the electron is the only particle produced in the interaction.
% Secondly, a separate analysis is considered for the channel in which the electron is produced in association with a certain number of protons (\nueccnopinp).
% This topology is mainly produced by charged current quasi elastic interactions (CCQE), interactions targeted by the MiniBooNE analysis: for this reason, an analysis targeting this channel is of primary importance.
% This channel is more powerful in rejecting the background, as the presence of additional protons allows an accurate reconstruction of the vertex.
% This channel is the main focus of the rest of the paper.
% Eventually, more complex events, with an arbitrary number of protons and additional hadrons (mainly pions) can be studied (\nueccmpinp).
% However, these events are typically result of deep inelastic scattering interactions with individual partons inside nucleons.
% Such events typically require larger neutrino energy, making them less relevant for a LEE search.
The second main search is for a photon-like excess, testing the hypothesis that the low energy excess is due to photons.
This second analysis exploits the typical separation between the main vertex and the beginning of the electromagnetic shower, as well as the energy deposition at the beginning of the shower corresponding to an $e^+e^-$ pair.
Further details about this analysis can be found in \cite{public_note_single_photon}.

\subsection{The \nueccnopinp analysis chain}

The \nueccnopinp analysis chain relies on three main steps.
First of all, in order to drastically remove the background produced by cosmic rays, compatibility between scintillation light recorded by the Photo Multiplier system and the charge detected in the TPC is required.
% Only events with a neutrino candidate well compatible with the flash that triggered the event, produced by the Argon scintillation light and recorded during the beam spill window, are selected.
Additionally, several cuts to ensure selection of good quality events and with the correct topology are applied.
% These contain fiducial volume requirements, and the presence of one well reconstructed shower and at least one well reconstructed track, with starting points compatible with the main vertex.
Eventually several cut to reject the background based on the calorimetry are applied.
% Among these, the identification of the proton track through the dE/dx versus residual range and identification of the shower through its energy deposition at the beginning are very powerful requirements for the selection.
Eventually the energy deposited in the TPC can be reconstructed by properly convert the charge to energy.
More details about this selection can be found in \cite{public_note_electron_neutrinos}.

\subsection{Sidebands and cross checks}
\begin{figure}
\begin{minipage}{0.5\linewidth}
\centerline{\includegraphics[width=0.95\linewidth]{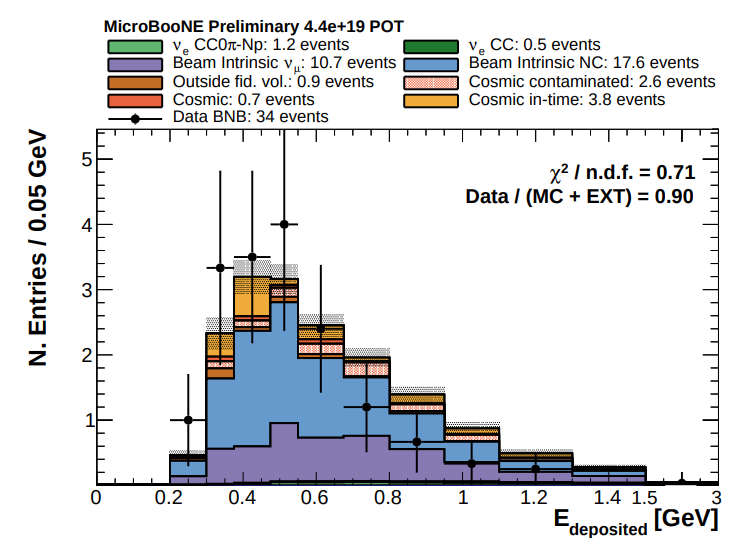}}
\end{minipage}
\hfill
\begin{minipage}{0.5\linewidth}
\centerline{\includegraphics[width=0.95\linewidth]{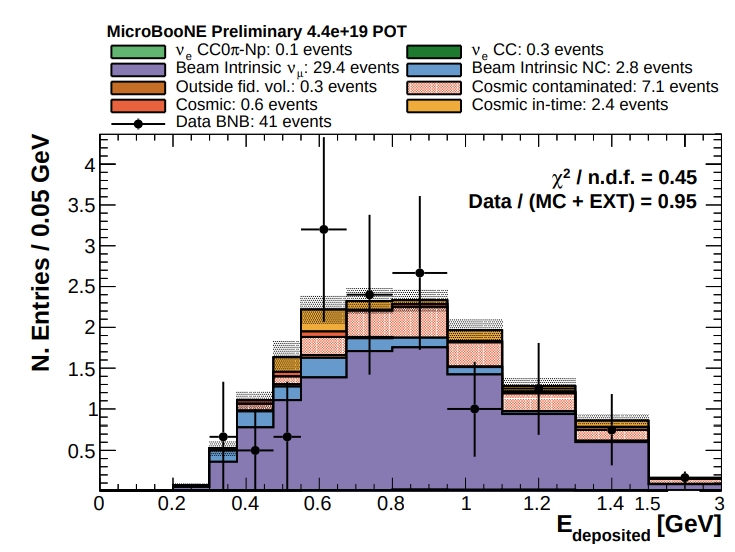}}
\end{minipage}
\caption[]{Distributions of the selected events in the NC $\pi^0$ sideband (left) and \numucc sideband (right) of the \nueccnopinp analysis.}
\label{fig:03_sidebands}
\end{figure}

The selection described in the previous section has been validated with two sidebands.
In the first one, the contribution of neutral current interactions with a production of a $\pi^0$ (NC $\pi^0$) is enhanced, by inverting the calorimetry requirements on the showers, in order to select photon-like showers produced by the $\pi^0$ decay.
Secondly, a sample enriched of \numucc interactions is selected by vetoing shower and requiring the presence of a long track compatible with a mip particle.
The distribution of the energy of the selected events in the two sidebands are shown in figure \ref{fig:03_sidebands}.
This result is also in good agreement with a more complete \numucc analysis, developed aiming at a cross section measurement \cite{public_note_muon_neutrinos}.
In both cases the agreement between the data and the simulation is satisfactory, making the analysis and the simulation reliable.
Eventually, this analysis has been reproduced on the NuMI dataset, which contains a larger number of \nuecc interaction per POT, within the same energy regime.
This test showed a good capability of the analysis in selecting \nueccnopinp interactions with a decent agreement between the data and the simulation.
A similar analysis, selecting \nuecc interactions inclusively also showed similar and encouraging result \cite{public_note_numi}.

\section{Conclusion and outlook}
A solid strategy for the Low Energy Excess search with the MicroBooNE detector, in the \nueccnopinp channel has been demonstrated, using different control regions and datasets.
The MicroBooNE collaboration is now preparing the final analyses, refining and improving the simulation as well as the reconstruction in order to achieve a larger efficiency for the selection of the signal and increase the chance to discover a possible low energy excess.

\bibliographystyle{ieeetr}

\end{document}